\documentclass[prl,twocolumn,preprintnumbers]{revtex4-2}%

\usepackage[T1]{fontenc}
\usepackage[a4paper]{geometry}
\usepackage{hyperref}
\usepackage{amsmath,amssymb,amsfonts}
\usepackage{comment}
\DeclareMathAlphabet\mathbfcal{OMS}{cmsy}{b}{n}

\usepackage{tikz}
\usepackage{tkz-euclide}
\usepackage{color}

\usetikzlibrary{decorations.markings}
\tikzset{middlearrow/.style={
		decoration={markings,
			mark= at position 0.5 with {\arrow{#1}} ,
		},
		postaction={decorate}
	}
}

\hypersetup{
  linktocpage,
  colorlinks  = true, 
  urlcolor    = blue, 
 linkcolor   = blue, 
  citecolor   = blue 
}

\def\be{\begin{equation}}
\def\ee{\end{equation}}

\definecolor{uuuuuu}{rgb}{0.26666666666666666,0.26666666666666666,0.26666666666666666}
\def\cbO{{\mathbfcal O}}
\def\cbL{{\mathbfcal L}}

\def \be  {\begin{equation}}
\def \ee  {\end{equation}}
\def \ba  {\begin{eqnarray}}
\def \ea  {\end{eqnarray}}

\def \Li	{ {\rm Li}}
\def \Ads2s2	{ AdS$_2\times$S$^2$ }

\def \cO	{ {\mathcal{O}}}

\def \cO{\mathcal{O}}

\begin{document}

\title{Hidden Conformal Symmetry in AdS$_2 \times$S$^2$ Beyond Tree Level}
\author{P. J. Heslop$^{a}$}
\author{A. E. Lipstein$^{a}$}
\author{M. Santagata$^{b}$ }
\affiliation{$^{a}$Department of Mathematical Sciences, Durham University, Durham, DH1 3LE, UK \\ $^b$Department of Physics, National Taiwan University, Taipei 10617, Taiwan}

\begin{abstract}
\noindent 
Correlators of $\mathcal{N}=2$ hypermultiplets with two-derivative interactions in AdS$_2\times$S$^2$ exhibit a hidden four-dimensional conformal symmetry which allows one to repackage all tree-level 4-point correlators into a single four-dimensional object corresponding to a contact diagram arising from a massless $\phi^4$ theory in AdS$_2\times$S$^2$. This theory serves as a toy model for IIB string theory in AdS$_5\times$S$^5$ and is interesting in its own right because AdS$_2\times$S$^2$ describes the near-horizon limit of extremal black holes in four dimensions. We argue that after acting with an $SU(1,1)\times SU(2)$ Casimir, all one-loop correlators can similarly be encoded by a four-dimensional function which arises from a one-loop scalar bubble diagram in AdS$_2\times$S$^2$, explaining how the hidden conformal symmetry extends beyond tree level. Finally, we conjecture a scalar effective field theory with a two-derivative interaction in AdS$_2\times$S$^2$ whose Witten diagrams should directly reproduce 4-point correlators to all loops without acting with Casimirs.
\end{abstract}

\maketitle

\section{Introduction}

The celebrated AdS/CFT correspondence \cite{Maldacena:1997re} relates quantum gravity in Anti-de Sitter (AdS) background to a non-gravitational conformal field theory (CFT) living in the boundary. The basic observables of a CFT are its correlation functions, which encode the scattering of gravitons and other degrees of freedom in AdS and can be computed perturbatively using Witten diagrams. For certain backgrounds of the form AdS$\times$S, where S is a sphere, simple formulae packaging tree-level correlation functions of all the modes on the sphere into a single generating function have been obtained, originating from  a hidden higher dimensional conformal symmetry. From the point of view of the dual CFT, these correspond to correlation functions of an infinite number of certain protected operators known as $1/2$-BPS operators.

Hidden conformal symmetry was first discovered in 4-point tree-level amplitudes  of type IIB supergravity in AdS$_5\times$S$^5$ \cite{Caron-Huot:2018kta}, which is dual to maximally supersymmetric Yang-Mills theory in four dimensions ($\mathcal{N}=4$ SYM) at strong coupling, and the symmetry has also been observed at weak coupling \cite{Caron-Huot:2021usw,Caron-Huot:2023wdh,Heslop:2022xgp}. If one includes higher-derivative corrections to supergravity, the higher dimensional conformal symmetry gets broken but there are still remnants of a higher dimensional organising principle \cite{Drummond:2019odu,Drummond:2020dwr,Aprile:2020luw,Aprile:2020mus}.
Indeed it was discovered that the four-point correlators of ${\mathcal N}=4$ SYM theory
arise from massless scalar contact Witten diagrams on AdS$_5 \times$S$^5$ \cite{Abl:2020dbx}. 

This higher dimensional symmetry was also found in AdS$_3\times$S$^3$ \cite{Rastelli:2019gtj,Giusto:2020neo}, AdS$_5 \times$S$^3$ \cite{Alday:2021odx} and AdS$_2 \times$S$^2$ \cite{Abl:2021mxo} backgrounds. The last case is of particular interest because the geometry arises in the near-horizon limit of four-dimensional extremal black holes \cite{Bertotti:1959pf,Robinson:1959ev} and may therefore have actual relevance for quantum gravity in the real world. The 4-point tree-level amplitudes of 4d $\mathcal{N}=2$ hypermultiplets interacting via two-derivative interactions in this background (which  arises from reducing $\mathcal{N} = 8$ supergravity on AdS$_2 \times$S$^2$ \cite{Michelson:1999kn,Lee:1999yu} and can be described as correlators of a 1d CFT with $SU(1,1|2)$ superconformal symmetry) can be encoded by a massless $\phi^4$ contact Witten diagram in the bulk \cite{Abl:2021mxo} very similarly to the IIB/ ${\mathcal N}=4$ SYM case. Note that a straightforward generalisation of this describes four-point open string scattering in IIB string theory on AdS$_5 \times$S$^5/\mathbb{Z}_2$ \cite{Glew:2023wik}. Hence the scalar theory in AdS$_2 \times$S$^2$ can be thought of as a toy model for IIB string theory on AdS$_5 \times$S$^5$. 

The purpose of this work is to go beyond the classical theory and investigate this at loop level. In particular, we consider four-point functions at one-loop order and show that they can be obtained by acting with Casimir operators on a 4d generating function. We then show that, quite remarkably, this generating function can be derived from a one-loop bubble diagram of the $\phi^4$ theory. We observe that the infinite sum over modes of the 2-sphere which flow through each edge of a Feynman diagram can be resummed into a single bulk-to-bulk propagator in AdS$_2 \times$S$^2$ which takes the same form as a propagator in four-dimensional flat space. 

Finally, we conjecture a full effective field theory in AdS$_2 \times$S$^2$ whose Feynman rules should directly compute four-point correlators in the two-derivative sector to all loop-orders without acting with Casimirs. The key insight that leads us to this action is that acting with a boundary Casimir on a contact diagram is equivalent to acting with a Laplacian in the bulk. We verify this proposal for free-theory and tree-level correlators and write down a one-loop formula whose explicit evaluation we leave for future work.

\section{basics of $SU(1,1|2)$ correlators}
The purpose of this section is to review what is currently known about four-point functions of half-BPS operators in the $SU(1,1|2)$ SCFT. For a pedagogical review of bootstrap techniques in superconformal field theory, see for example \cite{Heslop:2022xgp,Bissi:2022mrs}. We follow the conventions of \cite{Abl:2021mxo} and denote the $1/2$-BPS operators by $\mathcal{O}_p(x,y)$, where $x$ is the spacetime coordinate and $y$ an analogous internal coordinate dealing with the $SU(2)$ R-symmetry. These operators have protected dimension, and $SU(2)$ charge $p+\frac{1}{2}$ with $p=0,1,\ldots$. 

$\mathcal{O}_p$ is the primary operator of the superfield $O_p$:
\begin{equation}
O_p= \mathcal{O}_p+\theta \mathcal{L}_p +\bar{\theta} \bar{\mathcal{L}}_p+\cdots
\end{equation}
and so the descendant $\mathcal{L}_p$ has dimension $p{+}1$ and $SU(2)$ charge $p$.
It is extremely useful to combine operators for all $p$ into single operators $\cbO$, $\cbL$, $\bar{\cbL}$:
\begin{align}
	& \cbO= \sum_{p=0}^\infty \mathcal{O}_p\,,\qquad \cbL= \sum_{p=0}^\infty \mathcal{L}_p\,, \qquad \bar{\cbL}= \sum_{p=0}^\infty \bar{\mathcal{L}}_p\,.
\end{align}
Correlators of these objects will be referred to as ``master correlators'' and they act as 
generating functions in the sense that correlators with given external charges $p_i$  are obtained simply by Taylor-expanding in the $y$ variables and projecting onto the component with the correct conformal weight.

We will be interested in the four-point function of both the primaries and their descendants and we will expand  all correlators in a large central charge expansion, e.g.
\begin{equation}
	\langle \cbL \cbL \bar \cbL \bar \cbL \rangle =\sum_{m=0}^{\infty}\frac1{c^m}\langle \cbL \cbL \bar\cbL \bar\cbL \rangle^{(m)}\,,
\end{equation}
where the first term corresponds to free theory, the second to tree level amplitudes in the bulk, and so on.
The goal of this work is to compute the one-loop contribution $\langle \cbL \cbL \bar\cbL \bar\cbL \rangle^{(2)}$ and understand its structure.

Superconformal symmetry implies that the four-point function of  primaries and descendants are closely related, taking the form~\cite{Abl:2021mxo}
\begin{align}\label{LOrelation}
	\langle \cbO \cbO\cbO\cbO\rangle &= \langle \cbO \cbO\cbO\cbO\rangle^{(0)}  + I \times {\bf G}(x_i,y_i;c)\,,\notag\\
		\langle \cbL \cbL \bar\cbL \bar\cbL \rangle &=   \langle \cbL \cbL \bar\cbL \bar\cbL \rangle^{(0)} + \mathcal{C}_{12}    {\bf G} (x_i,y_i;c)\ .
\end{align}
Here $I$ is a certain kinematic factor given in (A4) of the Supplemental Material \cite{supplemental}, and $\mathcal{C}_{12}$ is the quadratic Casimir of $SU(1,1)\times SU(2)$ arising from the superconformal $SU(1,1|2)$ Casimir~\cite{Abl:2021mxo} applied at points 1 and 2:
\begin{align}\label{casimir}
	\mathcal{C}_{12}&=\mathcal{C}^x_{12}-\mathcal{C}^y_{12}\,,\notag\\
	\mathcal{C}^x_{12}&=\tfrac12\left(D_{x_1}{+}D_{x_2}\right)_{AB}\left(D_{x_1}{+}D_{x_2}\right)^{AB},
\end{align}
where
\begin{align}\label{generator}
	\qquad D_{xAB}=x_A \frac\partial{\partial{{ x}^B}}- x_B \frac\partial{\partial{{ x}^A}}
\end{align}
is a generator of the conformal or $SU(2)$ R-symmetry group. Note that we will mainly use 3-component embedding space coordinates $x^A,y^A$ throughout the paper. For more details on our conventions see appendix A and equation (D5) of the Supplemental Material \cite{supplemental}.
 
All the  correlators are functions of $x_i,y_i$, but if the hidden 4-dimensional symmetry is present then  the master correlators should be expressible in terms of functions of 4d distances, obtained by summing conformal and internal distances: 
\begin{equation}\label{4dinv}
\mathbf{x}_{ij}^2= x_{ij}^2+ y_{ij}^2\,,
\end{equation}
with the (boldface) 4d cross ratios defined accordingly:
\begin{align}
& \mathbf{u}=  \frac{\mathbf{x}_{12}^2\mathbf{x}_{34}^2 }{\mathbf{x}_{13}^2 \mathbf{x}_{24}^2 }=\mathbf{z}\mathbf{\bar{z}}\,,  \qquad  \mathbf{v}= \frac{\mathbf{x}_{14}^2\mathbf{x}_{23}^2 }{\mathbf{x}_{13}^2 \mathbf{x}_{24}^2 }=(1-\mathbf{z})(1-\mathbf{\bar{z}})\,.
\end{align}

Now we will illustrate how these variables play a role in free theory and at  tree level and then we will find that also loop level results can be written in terms of these variables albeit in a more non-trivial manner.

\vspace{0.5cm}

{\bf Free theory.} The 4d conformal  symmetry in free theory is realised at the level of correlators of descendants which can be written entirely in terms of 4d distances~\cite{Abl:2021mxo}
\begin{equation}\label{freetL}
\langle  \cbL\cbL \bar{\cbL} \bar{\cbL} \rangle ^{(0)} = \frac{1}{\mathbf{x}_{14}^2 \mathbf{x}_{23}^2}+ \frac{1}{ \mathbf{x}_{13}^2 \mathbf{x}_{24}^2}\, .
\end{equation}

{\bf Tree level.} 
At tree level on the other hand, the 4d conformal symmetry is directly realised on the correlators of primaries and we have
\begin{equation}\label{oooomaster}
{\bf G}^{(1)} = - D_{1111}(\mathbf{x}_{ij}^2),
\end{equation}
where $D_{1111}$ is the (normalised) one-loop box diagram:
\begin{equation}
\begin{split}
& D_{1111}(\mathbf{x}_{ij}^2) = \frac{1}{\mathbf{x}_{13}^2\mathbf{x}_{24}^2} \frac{\phi^{(1)} (\mathbf{z},\mathbf{\bar{z}})}{\mathbf{z}-\mathbf{\bar{z}}} =  \\
& = \frac{1}{\mathbf{x}_{13}^2\mathbf{x}_{24}^2}\frac{2\Big(\Li_2 (\mathbf{z}) - \Li_2 (\mathbf{\bar{z}})\Big) + \log \mathbf{u}  \log \left(\frac{1-\mathbf{z}}{1-\mathbf{\bar{z}}} \right) }{\mathbf{z}-\mathbf{\bar{z}}}\,.
\end{split}
\end{equation}
As mentioned before, the recipe to obtain the individual correlators is quite simple: one just expands the master correlator in the $y$ variables. For example, at the first order in the Taylor expansion, the master correlator yields the correlator with lowest charges:
\begin{equation}
G_{0000}^{(1)} = {\bf G}^{(1)}\vert_{y_i=0} = - \frac{1}{x_{12}^2  x_{34}^2} \Big( \frac{\log x^2}{1-x}+ \frac{\log (1-x)^2}{x} \Big)
\end{equation}
where $u=x^2=x_{12}^2x_{34}^2/(x_{13}^2x_{24}^2)$ is the 1d cross-ratio (recall that in 1d the cross ratios $u,v$ are not independent).
Note that the correlator, after Taylor expanding, is automatically written in 1d variables.

\section{One-loop correlators}

Let us now consider one-loop correlators. We will focus on the trascendental weight 2 part of the one-loop correlators, which can be uniquely fixed by requiring consistency of the double-discontinuity with the OPE and crossing symmetry.
Now, the double-discontinuity is simply related to the tree-level discontinuity via the action of the Casimir
\begin{equation}
\begin{split}
& {\bf G}^{(2)} \big|_{\log^2 u}  = - \frac{1}{2}  \mathcal{C}_{12}  (D_{1111}|_{\log \mathbf{x}_{12}^2})\,,
\end{split}
\end{equation}
where
\begin{equation}
(D_{1111})|_{\log  \mathbf{x}_{12}^2} = \frac{1}{\mathbf{x}_{13}^2\mathbf{x}_{24}^2( \mathbf{z}-\mathbf{\bar{z}})} \log  \left(\frac{1-\mathbf{z}}{1-\mathbf{\bar{z}}}\right)\,,
\end{equation}
The simplicity of this relation \cite{fn2} is really a consequence of the fact that the leading log is controlled by tree-level CFT data, which, in theories with hidden higher dimensional conformal symmetries, depend on the higher dimensional spin \cite{Aprile:2018efk,Caron-Huot:2018kta,Aprile:2021mvq,Abl:2021mxo,Drummond:2022dxd,Paul:2023zyr}. In this case the latter is truncated to spin 0 so there is only a single higher dimensional operator appearing in the OPE.

The crossing completion of the above double-discontinuity then automatically yields the complete result for the the maximal transcendental weight part of the one-loop correlator:
\begin{align}\label{generatorcorrelatoroneloop}
 {\bf G}^{(2)}  = -\frac{1}{2} &  \Big( 
	  \mathcal{C}_{12} \,\left(  (D_{1111})|_{\log \mathbf{x}_{12}^2} \right) \log u \log \frac u{v}  \notag \\ 
& + \mathcal{C}_{23}\, \left((D_{1111})|_{\log \mathbf{x}_{23}^2}\right) \log v \log \frac {v}u \notag \\ 
& +\mathcal{C}_{13} \, \left( (D_{1111})|_{\log \mathbf{x}_{13}^2} \right)\log u \log v \Big)\,,
\end{align}
where $\mathcal{C}_{12}$ is given in \eqref{casimir} and $\mathcal{C}_{23}$ and $\mathcal{C}_{13}$ are its crossing versions.
Note that the solution dictated by OPE and crossing symmetry carves out a three-dimensional space in the four-dimensional space of harmonic polylogarithms of weight 2, given by the three orientations of the logs. In particular, note the absence of $\text{Li}_2$.

\vspace{0.5cm}

{\bf A remarkable 4d uplift.}
Despite being quite simple, the one-loop formula presented above is still not very satisfactory from the point of view of the higher dimensional symmetry, because it is \emph{not} a function of 4d distances.
In fact, it turns out that there exists a function $\mathfrak{f}_s(\mathbf{z},\mathbf{\bar{z}})$ of four-dimensional distances \emph{only} such that
\begin{align}\label{generatorcorrelatoroneloopuplift}
	& {\bf G}^{(2)} = -\frac{1}{2}\Big( 
	  \mathcal{C}_{12} \frac{\mathfrak{f}_s(\mathbf{z},\mathbf{\bar{z}}) }{\mathbf{x}_{13}^2 \mathbf{x}_{24}^2}  + \mathcal{C}_{23} \frac{\mathfrak{f}_t(\mathbf{z},\mathbf{\bar{z}})}{\mathbf{x}_{13}^2 \mathbf{x}_{24}^2} +  \mathcal{C}_{13} \frac{\mathfrak{f}_u(\mathbf{z},\mathbf{\bar{z}}) }{\mathbf{x}_{13}^2 \mathbf{x}_{24}^2} \Big)\,,
\end{align} 
where
\begin{equation}\label{fsdef}
	\mathfrak{f}_s(\mathbf{z},\mathbf{\bar{z}}) = \frac{1}{3} \frac{1}{\mathbf{z}-\mathbf{\bar{z}}} \left( f^{(3)}(\mathbf{z},\mathbf{\bar{z}}) + \Big( \log \mathbf{u}  + \log \frac{\mathbf{u}}{\mathbf{v}} \Big)\phi^{(1)}(\mathbf{z},\mathbf{\bar{z}}) \right)\,,
\end{equation}
and
\begin{align}\label{ftfu}
& \mathfrak{f}_t (\mathbf{z},\mathbf{\bar{z}})=\mathfrak{f}_s (1-\mathbf{z},1-\mathbf{\bar{z}})\,, \qquad \mathfrak{f}_u (\mathbf{z},\mathbf{\bar{z}})= \frac{1}{\mathbf{z}\,\mathbf{\bar{z}}} \mathfrak{f}_s \Big(\frac{1}{\mathbf{z}},\frac{1}{\mathbf{\bar{z}}} \Big) \,.  \notag \\
\end{align}
Here $f^{(3)}$ is a weight-3 single-valued antisymmetric function which, in AdS context, recently made its first appearance in \cite{Drummond:2019hel}.

The function $\mathfrak{f}_s(\mathbf{z},\mathbf{\bar{z}})$
transforms correctly under $1\leftrightarrow 2$ crossing, and is a weight-3 combination of single-valued multiple polylogarithms (SVMPL's) \cite{Schnetz:2021ebf}. This is the expected weight for one-loop amplitudes in $d>1$ originating from a four-point contact interaction in the bulk \cite{Drummond:2019hel,Drummond:2020uni,Aprile:2022tzr,Paul:2023zyr,Heckelbacher:2022fbx}. In fact, $\mathfrak{f}_s$ is the \emph{unique} linear combination of SVMPL's built out of the alphabet $\{\mathbf{z},1{-}\mathbf{z},\mathbf{\bar{z}},1{-}\mathbf{\bar{z}},\mathbf{z}{-}\mathbf{\bar{z}} \}$  such that it correctly reproduces the double-discontinuity and, crucially, its crossing versions $\mathfrak{f}_t,\,\mathfrak{f}_u$ \emph{do not} contain $\log^2 \mathbf{u}$ terms.
We refer to appendix B of the Supplemental Material \cite{supplemental} for a more precise definition and a list of some useful properties.

We emphasize that the only assumption that went into the derivation of the one-loop formula for all higher charge correlators in \eqref{generatorcorrelatoroneloop} was the existence of higher dimensional conformal symmetry at tree level. The fact that it can be written in the form~\eqref{generatorcorrelatoroneloopuplift} is a very non-trivial consequence of this assumption.

\section{Generalised Witten diagrams}
One of the consequences of the existence of a hidden conformal symmetry is that tree-level correlators can be obtained from a massless $\phi^4$ contact Witten diagram in AdS$_2 \times$S$^2$ \cite{Abl:2021mxo}. The purpose of the remaining part of the paper is to test the applicability of such an approach  beyond tree level.
Before doing so, let us first recall how to obtain the tree level result from a $\phi^4$ generalised Witten diagram. For a pedagogical introduction to Witten diagrams in embedding space, see for example of \cite{Sleight:2016hyl}.

Points in AdS and S are parametrised in ($d{+}2$)-dimensional embedding space as  $-\hat{x}{\cdot}\hat x =\hat{y}{\cdot}\hat y= 1$, where we set the radii of AdS and S to one.
Analogously, boundary coordinates, denoted by un-hatted letters, satisfy $x{\cdot}x=y{\cdot}y=0$. 

\vspace{0.5cm}

{\bf Tree level.}
The following interaction term in the bulk effective action
\begin{equation}\label{bulktree}
S_{\text{int}} = \frac{1}{4!} \int_{{\rm AdS_2 \times S^2}} \, d^2 \hat x_0 d^2 \hat y_0 \, \phi (\hat x_0,\hat y_0)^4
\end{equation}
computes all tree-level correlators of the supersymmetric CFT using generalised Witten diagrams which treat AdS and S on equal footing. 
In particular all half-BPS correlators are generated by the following AdS$_2\times$S$^2$ integral, which is consistent with the bootstrapped result~\eqref{oooomaster}:
\begin{equation}\label{ooootree}
\begin{split}
{\bf G}^{(1)} &  = - \frac{1}{\pi^2}   \int \frac{d^{2}\hat{x_0} d^{2} \hat{y_0}}{\mathbf{x}^2_{10}\mathbf{x}^2_{20}\mathbf{x}^2_{30}\mathbf{x}^2_{40}}=  - D_{1111}(\mathbf{x}_{ij}^2),
\end{split}
\end{equation}
where $
\mathbf{x}^2_{i0} = -2x_{i}{\cdot}\hat{x}_0-2 y_{i}{\cdot}\hat y_0$
is the (inverse of the) AdS$\times$S bulk-to-boundary propagator, which satisfies 
\begin{equation}
\nabla^2\frac{1}{\mathbf{x}^2_{i0}}=\left(\nabla_{\hat{x}_0}^2+\nabla_{\hat{y}_0}^2\right)\frac{1}{\mathbf{x}^2_{i0}}=0\,,
\label{eom}
\end{equation}
where  $\nabla_{\hat{x}_0}^2$ ($\nabla_{\hat{y}_0}^2$) is the AdS (S) Laplacian, which is given in equation \eqref{bulklaplacian} below.
We refer to \cite{Abl:2021mxo} for a detailed derivation of \eqref{ooootree}.

\vspace{0.5cm}

{\bf One loop.}
In order to test whether the effective action extends beyond tree level, we first need to find an expression for the AdS$_2\times$S$^2$ bulk-to-bulk propagator.
Intriguingly, it turns out that  the bulk-to-bulk propagator in the  AdS$_{d+1}\times$S$^{d+1}$ product geometry takes exactly the same functional form as the bulk-to-boundary propagator: 
\begin{align}\label{bulktobulk}
\frac1{{\bf \hat x}^{2d}_{lr}}= \frac1{(-2\hat x_{l}{\cdot}\hat x_{r} -2 \hat y_{l}{\cdot}\hat y_{r})^d}\ .
\end{align}
In fact, in AdS$\times$S this simple propagator satisfies \eqref{eom}.  This is very different from pure AdS space where bulk-to-bulk propagators are hypergeometric functions. Remarkably, this propagator also encodes an infinite sum over the Kaluza-Klein modes of the sphere propagating in AdS:
\begin{equation}\label{bulktobulksumm}
\begin{split}
& \frac1{{\bf \hat x}^{2d}_{lr}}= \frac{(-2)^{-d}}{(d-1)!} \sum_{p=0}^\infty  \Big[ (-1)^{\tilde{p}} (p-d+1)_{d-1} \\ 
& \times \frac{1}{(\hat{x}_l {\cdot} \hat{x}_r)^p} \, {}_2F_1 \Big (\frac{p{+}1}{2},\frac{p}{2},p{+}1{-}\frac{d}{2};\frac{1}{ (\hat{x}_l{\cdot} \hat{x}_r)^2} \Big)  \\
&\times \frac{1}{(\hat{y}_l {\cdot} \hat{y}_r)^{\tilde{p}}} \, {}_2F_1 \Big (\frac{\tilde{p}{+}1}{2},\frac{\tilde{p}}{2},\tilde{p}{+}1{-}\frac{d}{2};\frac{1}{(\hat{y}_l{\cdot} \hat{y}_r)^2}\Big) \Big]\,,
 \end{split}
\end{equation}
where $\tilde{p}=d-p$. Similar structures were derived for IIB supergravity in AdS$_5\times$S$^5$ in \cite{Dai:2009zg}. 

With the bulk-to-bulk propagator at hand, we can then compute a one-loop bubble diagram in AdS$_{d+1}\times$S$^{d+1}$ which in the s-channel reads (up to an unimportant overall constant)
\begin{equation}\label{ks}
\begin{tikzpicture}[x=.1cm,y=.1cm, anchor=base, baseline,label distance=.5mm,inner sep=0mm]
	\tikzstyle{vertex}=[fill=black, draw=black, shape=circle,inner sep=0pt,minimum size=3]
	\tikzstyle{none}=[]
	\node [style=none, label={left:$\scriptscriptstyle 1$}] (0) at (-4, 3) {};
	\node [style=none, label={left:$\scriptscriptstyle 2$}] (1) at (-4, -3) {};
	\node [style=none, label={right:$\scriptscriptstyle 3$}] (2) at (4, 3) {};
	\node [style=none, label={right:$\scriptscriptstyle 4$}] (3) at (4, -3) {};
	\node [style=vertex, label={below:$\scriptscriptstyle r$}] (4) at (2, 0) {};
	\node [style=vertex, label={below:$\scriptscriptstyle l$}] (5) at (-2, 0) {};
	\draw [line width=.06] (0.center) to (5.center);
	\draw [line width=.06] (1.center) to (5.center);
	\draw [line width=.06] (4.center) to (2.center);
	\draw [line width=.06] (4.center) to (3.center);
	\draw [line width=.06] [in=135, out=45, looseness=1.25] (5.center) to (4.center);
	\draw [line width=.06] [bend right=45, looseness=1.25] (5.center) to (4.center);
	\draw (0,0) circle (5);
\end{tikzpicture}
\propto \mathcal{K}_{s}\equiv\int\frac{d^{d+1}\hat{x}_{l}d^{d+1}\hat{x}_{r}d^{d+1}\hat{y}_{l}d^{d+1}\hat{y}_{r}}{\mathbf{x}_{1l}^{2d}\mathbf{x}_{2l}^{2d}\mathbf{x}_{3r}^{2d}\mathbf{x}_{4r}^{2d}\hat{\mathbf{x}}_{lr}^{4d}}\,,
\end{equation}
with a similar formula for the $t$- and $u$-channels.

The double integral over the sphere can be evaluated combinatorially for arbitrary $d$ and $p_i$, see appendix C of the Supplemental Material \cite{supplemental} for more details.
Here, we will focus on $d=1,p_i=0$, for which the integral reduces to
\begin{equation}\label{Ksp=1}
\mathcal{K}_{s}\big|_{\substack{d=1\\y_{i}=0}}=\!\!\int\!\!\frac{d^{2}\hat{x}_{l}d^{2}\hat{x}_{r}\times 2^{-6}}{(x_{1}{\cdot} \hat{x}_l)(x_{2}{\cdot}\hat{x}_l)(x_{3}{\cdot}\hat{x}_r)(x_{4}{\cdot}\hat{x}_r)\big((\hat{x}_{l}{\cdot}\hat{x}_{r})^{2}{-}1\big)}\,.
\end{equation}
The remaining integration over  AdS produces divergences and needs to be regularised. This can be done, for example, following the procedure outlined in \cite{Heckelbacher:2022fbx}. In short, the idea is that the integral can be mapped to a flat-space integral so that one can take advantage of standard flat space techniques, such as dimensional regularisation, to isolate the divergence and extract the finite part. We refer to appendix D of the Supplemental Material \cite{supplemental} for details on the computation.
Here we just quote the result for the finite part, which reads
\begin{equation}
\mathcal{K}_s^{(\text{fin})} \big|_{\substack{d=1\\y_{i}=0}} \propto 
\left.
\frac{\mathfrak{f}_s (\mathbf{z},\mathbf{\bar{z}})  }{\mathbf{x}_{13}^2 \mathbf{x}_{24}^2}\
\right|_{y_i=0}
\label{bubble2}
\end{equation}
and analogously for $t$- and $u$-channels.
Quite remarkably, this is precisely the function in equation~\eqref{generatorcorrelatoroneloopuplift} appearing in the master correlator of the $SU(1,1|2)$ CFT!
The computation of the bubble diagram beyond $y_i=0$ is much more complicated and we leave it to the future but it is natural to expect that the relation~\eqref{bubble2} will extend beyond the $y_i=0$ case.

At this point, it is important to remark on the difference compared to tree-level: the bubble diagram at one loop  reproduces the correlator only upon acting with a Casimir in each crossing orientation~\eqref{generatorcorrelatoroneloopuplift}.
This leads to a natural question: is there a way to explain the presence of the Casimir from a bulk perspective and make a more precise statement? We will address this in the next section.

\section{An all-loops conjecture} \label{sec:eftdesc}
The key observation we will make use of is that acting with a boundary Casimir on a pair or bulk-to-boundary propagators is equivalent to acting with the AdS${}\times$S Laplacian in the bulk. This is a generalisation of similar observations in AdS~\cite{Eberhardt:2020ewh,Herderschee:2022ntr}.
The AdS and S Laplacians are equal (up to an important sign)  to the $SO(d,2)$ or $SO(d+2)$ Casimir operators acting on bulk coordinates:
\begin{align}\label{bulklaplacian}
	\nabla^2_{\hat x}=-\tfrac12 D_{\hat xAB}D_{\hat x}^{AB}\qquad 
	\nabla^2_{\hat y}=+\tfrac12 D_{\hat yAB}D_{\hat y}^{AB},\ 
\end{align}
where $D_{\hat xAB}$ is a generator, identical in form to the conformal generator on the boundary~\eqref{generator}.
When the bulk Laplacian acts on a pair of bulk-to-boundary propagators meeting at common point in the bulk, it can be traded with a Casimir acting on the boundary points: 
\begin{equation}\label{relCasLap}
 \nabla^2\left(\frac{1}{\mathbf{x}^2_{10} \mathbf{x}^2_{20}} \right) =- \mathcal{C}_{12}\left(\frac{1}{\mathbf{x}^2_{10} \mathbf{x}^2_{20}}\right) \,,
\end{equation}
where $\nabla^2=\nabla^2_{\hat x}+\nabla^2_{\hat y}$ is the AdS$\times$S Laplacian, and $\mathcal{C}_{12}=\mathcal{C}^x_{12}-\mathcal{C}^y_{12}$ is the  AdS$_2\times$S$^2$ Casimir~\eqref{casimir}.

Now note that the master correlator of descendants $\langle \cbL \cbL \bar\cbL \bar\cbL \rangle$ enjoys the following properties: in free theory the higher dimensional symmetry is manifest~\eqref{freetL}; at tree level it takes the form of the Casimir $\mathcal{C}_{12}$ acting on an object with higher dimensional symmetry~(\ref{LOrelation}, \ref{oooomaster}); at one loop it can be written as the Casimir $\mathcal{C}_{12}$ acting on further Casimirs which themselves act on functions with higher dimensional symmetry~(\ref{LOrelation}, \ref{generatorcorrelatoroneloopuplift}).
Along with the observation that external Casimirs can arise from a bulk Laplacian operator, this suggests the following effective action AdS$\times$S space (including kinetic terms):
\begin{equation}\label{llbarction}
S' =\int_{{\rm AdS_2 \times S^2}} d^{4}\hat{\bf x}_{0}  \left( \tfrac12 \bar\phi \nabla^2 \phi   -  \tfrac{G_N}{4}  \, \bar{\phi}^2  \nabla^2 \phi^2 \right),
\end{equation}
where we identify the \emph{complex} field $\phi$ as the bulk field coupled to the descendant operator $\cbL$. Here the Newton's constant is proportional to the inverse of the central charge, $G_N\propto 1/c$ and we employed the notation $d^{4}\hat{\bf x}_{0}\equiv d^{2}\hat{ x}_{0}d^{2}\hat{ y}_{0}$.

Th effective action \eqref{llbarction} is conjectured to describe all-loop four-point correlators of $\mathcal{N}=2$ hypermultiplets which arise from dimensional reduction of $\mathcal{N}=8$ supergravity on AdS$_2 \times$S$^2$. The effective action is therefore non-renormalisable, but it is straightforward to include higher derivative corrections which describe the low energy expansion of a UV completion like string or M-theory. A crucial property of $\frac{1}{2}$-BPS 4-point correlators which allows them to be described by a scalar effective action in the bulk is the ability to factor out a polynomial which encodes all the supersymmetry. A similar property holds for certain higher-point correlators known as maximally nilpotent correlators \cite{Eden:1999gh,Howe:1999hz,Eden:2011we}, so our effective action may also be applicable at higher points, although we leave an exploration of this possibility to future work. Let us now work out the predictions for 4-point correlators arising from this action. 

\vspace{0.5cm}

{\bf Free theory.}
By performing Wick contraction, the free-theory four-point function reads
\begin{equation}
\begin{split}
\langle \cbL \cbL \bar{\cbL}  \bar{\cbL} \rangle^{(0)}  & =	\begin{tikzpicture}[x=.1cm,y=.1cm,anchor=base, baseline,label distance=.5mm,inner sep=0mm]
	\tikzstyle{vertex}=[fill=black, draw=black, shape=circle]
	\tikzstyle{none}=[]
	\node [style=none, label={left:$ \scriptstyle 1$}] (0) at (-4, 3) {};
	\node [style=none, label={left:$ \scriptstyle 2$}] (1) at (-4, -3) {};
	\node [style=none, label={right:$\scriptstyle 3$}] (2) at (4, 3) {};
	\node [style=none, label={right:$\scriptstyle 4$}] (3) at (4, -3) {};
	\draw [middlearrow={latex}, line width=.1mm] (0.center) to (2.center);
	\draw [middlearrow={latex}, line width=.1mm] (1.center) to (3.center);
	\draw (0,0) circle (5);
\end{tikzpicture}\ +\ 
\begin{tikzpicture}[x=.1cm,y=.1cm,anchor=base, baseline,label distance=.5mm,inner sep=0mm]
	\tikzstyle{vertex}=[fill=black, draw=black, shape=circle]
	\tikzstyle{none}=[]
	\node [style=none, label={left:$\scriptstyle  1$}] (0) at (-4, 3) {};
	\node [style=none, label={left:$ \scriptstyle 2$}] (1) at (-4, -3) {};
	\node [style=none, label={right:$\scriptstyle 3$}] (2) at (4, 3) {};
	\node [style=none, label={right:$\scriptstyle 4$}] (3) at (4, -3) {};
	\draw [middlearrow={latex}, line width=.1mm] (0.center) to (3.center);
	\draw [middlearrow={latex}, line width=.1mm] (1.center) to (2.center);
	\draw (0,0) circle (5);
\end{tikzpicture} \\
& = \frac{1}{\mathbf{x}_{13}^2 \mathbf{x}_{24}^2}+ \frac{1}{ \mathbf{x}_{14}^2 \mathbf{x}_{23}^2}\,,
\end{split}
\end{equation}
which indeed reproduces the result for the free theory correlator \cite{Abl:2021mxo}, quoted in \eqref{freetL}.

\vspace{0.5cm}

{\bf Tree level.}
Following the standard AdS/CFT procedure, we can readily find the tree level four-point function, which reads:
\begin{equation}
\begin{split}
\langle \cbL \cbL \bar{\cbL}  \bar{\cbL} \rangle^{(1)} & =
\begin{tikzpicture}[x=.15cm,y=.15cm, anchor=base, baseline,label distance=.5mm,inner sep=0mm]
	\tikzstyle{vertex}=[fill=black, draw=black, shape=circle,inner sep=0pt,minimum size=5]
	\node [label={left:$\scriptstyle 1$}] (0) at (-4, 3) {};
	\node [label={left:$\scriptstyle  2$}] (1) at (-4, -3) {};
	\node [label={right:$\scriptstyle 3$}] (2) at (4, 3) {};
	\node [label={right:$\scriptstyle 4$}] (3) at (4, -3) {};
	\node [style=vertex] (4) at (0, 0) {};
		\draw [middlearrow={latex}, line width=.1mm] (0.center) to (4.center);
	\draw [middlearrow={latex}, line width=.1mm] (1.center) to (4.center);
	\draw [middlearrow={latex}, line width=.1mm] (4.center) to (2.center);
	\draw [middlearrow={latex}, line width=.1mm] (4.center) to (3.center);
	\draw (0,0) circle (5);
\end{tikzpicture}\\
&=
\frac{1}{\pi^2}\int d^{4}\hat{\bf x}_{0} \nabla^2 \Big(\frac{1}{\mathbf{x}^2_{10}\mathbf{x}^2_{20}}\Big) \frac{1}{\mathbf{x}^2_{30}\mathbf{x}^2_{40}}   \\
& =\frac{-1}{\pi^2} \mathcal{C}_{12}  \int \frac{ d^{4}\hat{\bf x}_{0} }{\mathbf{x}^2_{10}\mathbf{x}^2_{20}\mathbf{x}^2_{30}\mathbf{x}^2_{40}}  = - \mathcal{C}_{12} D_{1111}(\mathbf{x}_{ij}^2)
\,.
\end{split}
\end{equation}
The arrows on the vertex distinguish the pairs of legs on which the Laplacian acts.
This indeed reproduces the result for the master correlator~(\ref{LOrelation}, \ref{oooomaster}).

\vspace{0.5cm}

{\bf One loop.}
Finally, let us consider the one-loop amplitude.
Here we have to consider three orientations of the bubble diagram

\begin{align*}		
 \langle \cbL \cbL \bar{\cbL}  \bar{\cbL} \rangle^{(2)} = \begin{tikzpicture}[x=.14cm,y=.14cm,anchor=base, baseline,label distance=.5mm,inner sep=0mm]
		\tikzstyle{vertex}=[fill=black, draw=black, shape=circle,inner sep=0pt,minimum size=5]
		\tikzstyle{none}=[]
		\node [style=none, label={left:$\scriptstyle 1$}] (0) at (-4, 3) {};
		\node [style=none, label={left:$\scriptstyle 2$}] (1) at (-4, -3) {};
		\node [style=none, label={right:$\scriptstyle 3$}] (2) at (4, 3) {};
		\node [style=none, label={right:$\scriptstyle 4$}] (3) at (4, -3) {};
		\node [style=vertex, label={below:$\scriptstyle r$}] (4) at (2, 0) {};
		\node [style=vertex, label={below:$\scriptstyle l$}] (5) at (-2, 0) {};
		\draw [middlearrow={latex}, line width=.06] (0.center) to (5.center);
		\draw [middlearrow={latex}, line width=.06] (1.center) to (5.center);
		\draw [middlearrow={latex}, line width=.06] (4.center) to (2.center);
		\draw [middlearrow={latex}, line width=.06] (4.center) to (3.center);
		\draw [middlearrow={latex}, line width=.06] [in=135, out=45, looseness=1.25] (5.center) to (4.center);
		\draw [middlearrow={latex}, line width=.06] [bend right=45, looseness=1.25] (5.center) to (4.center);
		\draw (0,0) circle (5);
	\end{tikzpicture}
+
\begin{tikzpicture}[x=.14cm,y=.14cm,anchor=base, baseline,label distance=.5mm,inner sep=0mm]
	\tikzstyle{vertex}=[fill=black, draw=black, shape=circle,inner sep=0pt,minimum size=5]
	\tikzstyle{none}=[]
	\node [style=none, label={right:$\scriptstyle 3$}] (0) at (4, 3) {};
	\node [style=none, label={left:$\scriptstyle 1$}] (1) at (-4, 3) {};
	\node [style=none, label={right:$\scriptstyle 4$}] (2) at (4, -3) {};
	\node [style=none, label={left:$\scriptstyle 2$}] (3) at (-4, -3) {};
	\node [style=vertex, label={below:$\scriptstyle r$}] (4) at (0, -2) {};
	\node [style=vertex, label={above:$\scriptstyle l$}] (5) at (0, 2) {};
	\draw [middlearrow={latex}, line width=.06] (5.center) to (0.center);
	\draw [middlearrow={latex}, line width=.06] (1.center) to (5.center);
	\draw [middlearrow={latex}, line width=.06] (4.center) to (2.center);
	\draw [middlearrow={latex}, line width=.06] (3.center) to (4.center);
	\draw [middlearrow={latex}, line width=.06] [in=60, out=-60, looseness=1.25] (5) to (4);
	\draw [middlearrow={latex}, line width=.06] [bend right=-30, looseness=1.25] (4) to (5);
	\draw (0,0) circle (5);
\end{tikzpicture}
+
\begin{tikzpicture}[x=.14cm,y=.14cm,anchor=base, baseline,label distance=.5mm,inner sep=0mm]
	\tikzstyle{vertex}=[fill=black, draw=black, shape=circle,inner sep=0pt,minimum size=5]
	\tikzstyle{none}=[]
	\node [style=none, label={right:$\scriptstyle 3$}] (0) at (4.56, 2.04) {};
	\node [style=none, label={left:$\scriptstyle 1$}] (1) at (-4, 3) {};
	\node [style=none, label={right:$\scriptstyle 4$}] (2) at (4.56, -2.04) {};
	\node [style=none, label={left:$\scriptstyle 2$}] (3) at (-4, -3) {};
	\node [style=vertex, label={below:$\scriptstyle r$}] (4) at (0, -2) {};
	\node [style=vertex, label={above:$\scriptstyle l$}] (5) at (0, 2) {};
	\draw [middlearrow={latex}, line width=.06] (5.center) to (2.center);
	\draw [middlearrow={latex}, line width=.06] (1.center) to (5.center);
	\draw [middlearrow={latex}, line width=.06] (4.center) to (0.center);
	\draw [middlearrow={latex}, line width=.06] (3.center) to (4.center);
	\draw [middlearrow={latex}, line width=.06] [in=60, out=-60, looseness=1.25] (5) to (4);
	\draw [middlearrow={latex}, line width=.06] [bend right=-30, looseness=1.25] (4) to (5);
	\draw (0,0) circle (5);
\end{tikzpicture} \,.
\end{align*}
These yield the corresponding expressions
\begin{align}&\int d^{4}\hat{\bf x}_l d^{4}\hat{\bf x}_r\,\, \nabla_l^2\left(\frac1{{\bf x}^2_{1l}}\frac1{{\bf x}^2_{2l}}\right) \frac1{\widehat{\mathbf{x}}_{lr}^4}\nabla_r^2\left(\frac1{{\bf x}^2_{3r}} \frac1{{\bf x}^2_{4r}}\right)\,,	\notag\\
& \int d^{4}\hat{{\bf x}}_l d^{4}\hat{{\bf x}}_r\,\, \nabla_l^2\left(\frac1{{\bf x}^2_{1l}}\frac1{\widehat{\mathbf{x}}^2_{lr}}\right) \frac1{{\bf x}^2_{3l}{\bf x}^2_{4r}}\nabla_r^2\left(\frac1{{\bf x}^2_{2r}\widehat{\mathbf{x}}^2_{lr}} \right)\,,
\end{align}
with the $u$-channel expression obtained by swapping $(x_3,y_3) \leftrightarrow (x_4,y_4)$ in the $t$-channel expression.

The first expression is directly related (via \eqref{relCasLap}) to the bubble integral evaluated in the previous section:
\begin{align}
& \int d^{4}\hat{\bf x}_l d^{4}\hat{\bf x}_r\,\, \nabla_l^2\left(\frac1{{\bf x}^2_{1l}}\frac1{{\bf x}^2_{2l}}\right) \frac1{\widehat{\mathbf{x}}_{lr}^4}\nabla_r^2\left(\frac1{{\bf x}^2_{3r}} \frac1{{\bf x}^2_{4r}}\right)\, \notag\\
& = \mathcal{C}_{12}^2 \mathcal{K}_s |_{d=1},
\end{align}
and thus it correctly yields the $s$-channel part of the correlator~(\ref{LOrelation}, \ref{generatorcorrelatoroneloopuplift}) (at least for the $y_i=0$ component).
The evaluation of the integral in the other two channels is unfortunately less trivial and we will reserve it for future studies.

\section{Conclusion}

The study of quantum gravity in curved background is both technically and conceptually very challenging, but great progress has been made in AdS background because in this case there is a dual description in terms of a conformal field theory in the boundary whose correlation functions are strongly constrained by conformal and crossing symmetry. In recent years a new symmetry principle has emerged for certain AdS$\times$S backgrounds known as hidden conformal symmetry. In this paper, we explored the implications of this symmetry for 4-point correlators of $\mathcal{N}=2$ hypermultiplets arising from dimensional reduction of $\mathcal{N}=8$ supergravity on AdS$_2\times$S$^2$ at the quantum level. 

This paper has two main results: first, we found that the weight-2 transcendental sector at one loop can be organised into a 4d master correlator, which precisely coincides with the $y_i=0$ component of the one-loop bubble diagram of a massless $\phi^4$ theory in AdS$_2\times$S$^2$, upon acting with a Casimir in each orientation of the bubble. Moreover, we conjectured a scalar effective field theory in AdS$_2\times$S$^2$ which gives rise to the aforementioned Casimir via a two-derivative interaction in the bulk, and should produce all 4-point correlators to all loops in the two-derivative sector. It would be interesting to include higher derivative interactions to the effective action in \eqref{llbarction}, which would arise from the low energy expansion of a generic UV completion and work out their implications for loop-level correlators.

In this paper we have only dealt with the maximal transcendental weight part of the correlators. It would be interesting to see whether the generating function \eqref{generatorcorrelatoroneloopuplift} automatically incorporates the full structure dictated by the OPE. Moreover, it would be very useful to test our conjectured action in \eqref{llbarction} for correlators with arbitrary charges as well as for higher loops, and to understand how the structures we found in this paper generalise to higher-dimensional theories. For example, does a one-loop massless scalar bubble diagram in AdS$_5 \times$S$^5$ have anything to do with correlators in $\mathcal{N}=4$ SYM? In fact, while a (Mellin) formula for arbitrary charges at order $\lambda^{-\frac{3}{2}}$ is known \cite{Aprile:2022tzr}, it is still not clear if and how this can be written in terms of a generating function, which will most likely involve the sixth order differential operator found in \cite{Abl:2021mxo}.

While the correlators we considered have superconformal symmetry, the bulk effective theories we constructed are non-supersymmetric because it is possible to factor out a polynomial which encodes all the supersymmetry from the four-point correlators, so it is conceivable that the these effective can be adapted to describe less idealised situations. Indeed, since AdS$_2\times$S$^2$ arises in the near horizon limit of extremal black holes, the approach developed in this paper may ultimately provide powerful new tools for the study of semi-realistic black holes using effective field theory. Moreover, 4-point correlators in asymptotically AdS backgrounds are known to encode bound-states, quasi-normal modes, and gravitational radiation \cite{Dodelson:2022eiz}. It would therefore be very exciting if these ideas can be combined with the ones developed in this paper to compute real-world gravitational wave observables.

\section*{Acknowledgments}
\begin{acknowledgments}
We thank James Drummond, Hynek Paul, Ivo Sachs, and Pierre Vanhove for useful discussions. AL and PH are supported by an STFC Consolidated Grant ST/T000708/1. MS is supported by the Ministry of Science and Technology (MOST) through the grant 110-2112-M-002-006-.
\end{acknowledgments}

\hfill \break

\appendix

\section{Internal coordinate conventions} \label{conventions}

The formulae in this paper are all written in terms of $x,y$ coordinates in embedding space (either for the boundary or bulk spaces) where they have three components. The operators $\cO_p$ are however defined in terms of the 2-component $y_a$, homogeneous coordinates of the Riemann sphere $ \mathbb{CP}^1$. The two are related as follows
\begin{align}
    y_ay_b = y_A ( \sigma^A \epsilon )_{ab}\ .
\end{align}
with $\sigma^A$ Pauli matrices.
Choosing coordinates $y^a=(1,y)$ this gives 
\begin{align}
    y^A=\Big( \tfrac12{(y^2{-}1)}, -\tfrac i2 ({y^2{+}1}),y\Big)
\end{align}
which satisfies $y{\cdot}y=0$. 
Then we define the invariant combinations of two of these
\begin{align}\label{ys}
    y_{ij}^2:=(y_i{-}y_j){\cdot}(y_i{-}y_j)=-2y_i{\cdot}y_j=(y_i{-}y_j)^2
\end{align}
where the last expression is given in terms of the 1d Poincare patch $y_i$ coordinates but has exactly the same form as the second expression where the dot indicates the dot product in 3d embedding space.

The bulk $\hat y$'s (we use the terms ``boundary'' and ``bulk'' for the internal coordinates by analogy with AdS even though a sphere has no boundary) for the $\text{S}^2$ sphere are also given as 3-component vectors satisfying $\hat y{\cdot} \hat y=1$ (we assume  a unit radius for the sphere and AdS space). 

A completely  analogous story to the above applies for the  $x,\hat x$ coordinates where the dot product assumes the $(-,-,+)$ metric and $\hat x {\cdot} \hat x=-1$. More details can be found in Appendix~\ref{adsdetails}.

In these conventions, the expression for  the kinematic factor $I$ in~\eqref{LOrelation} is given as 
\begin{align}\label{II}
     I=x_{12}x_{34}y_{13}y_{24}-y_{12}y_{34}x_{13}x_{24}\ .
\end{align}
This is most straightforwardly viewed in the Poincare patch with $y_{ij}:=y_i{-}y_j,\ x_{ij}:=x_i{-}x_j$, but could equally be viewed in embedding space with $y_{ij}:=\sqrt{(y_i{-}y_j){\cdot}(y_i{-}y_j)},\ x_{ij}:=\sqrt{(x_i{-}x_j){\cdot}(x_i{-}x_j)}$\ .

Note that the  4d invariants defining boundary-to-boundary, bulk-to-boundary and bulk-to-bulk propagators  are written~\eqref{4dinv}
\begin{align}
    {\bf x}^2_{ij}&\!:=\!(x_{i}-x_j){\cdot}(x_i{-}x_j){+}(y_{i}{-}y_j){\cdot}(y_i{-}y_j)\!=\!{-}2x_i{\cdot}x_j{-}2y_i{\cdot}y_j\notag\\
    {\bf x}^2_{i0}&\!:=\!(x_{i}{-}\hat x_0){\cdot}(x_i{-}\hat x_0){+}(y_{i}{-}\hat y_0){\cdot}(y_i{-}\hat y_0)\!=\!{-}2x_i{\cdot}\hat x_0{-}2y_i{\cdot}\hat y_0\notag\\
 \hat{\bf x}^2_{lr}&\!:=\!(\hat x_{l}{-}\hat x_r){\cdot}(\hat x_l{-}\hat x_r){+}(\hat y_{l}{-}\hat y_r){\cdot}(\hat y_l{-}\hat y_r)\!=\!{-}2\hat 
 x_l{\cdot}\hat x_r{-}2\hat y_l{\cdot}\hat y_r 
\end{align}
with the equality on the right due to the cancellations of the form $\hat x{\cdot}\hat x+\hat y{\cdot}\hat y=-1+1=0$.

\section{The functions $f^{(3)}$ and $\mathfrak{f}_s$} \label{f3}
In this Appendix we will give more details about the weight-3 function appearing in the one-loop correlators. The function $f^{(3)}$ can be simply defined in terms of its total derivative:
\begin{align}\label{f3def}
  d f^{(3)} =\,\,\,&\Big(4 \phi^{(1)} -\log^2 \mathbf{v}+2\log \mathbf{u} \log \mathbf{v} \Big) d \log \mathbf{z}\, \notag  \\ 
 + & \Big(4\phi^{(1)}+\log^2 \mathbf{v}-2\log \mathbf{u} \log \mathbf{v} \Big) d \log \mathbf{\bar{z}}\, \notag  \\
 + & \Big(4\phi^{(1)}+\log^2 \mathbf{u}-2\log \mathbf{u} \log \mathbf{v} \Big) d \log (1-\mathbf{z})\,  \notag  \\
+ & \Big(4\phi^{(1)}-\log^2 \mathbf{u}+2\log \mathbf{u} \log \mathbf{v} \Big) d \log (1-\mathbf{\bar{z}})\, \notag  \\
 + & \Big(-12\phi^{(1)} \Big) d\log(\mathbf{z}-\mathbf{\bar{z}})\,.
\end{align}
The interested reader can find the full integrated expression -- given in terms of SVMPL's -- in \cite{Drummond:2019hel},\footnote{The expression here differs by an overall factor of -2 compared to the definition given in \cite{Drummond:2019hel}.}.
Note that $f^{(3)}$ satisfies
\begin{equation}\label{crossingf3}
 f^{(3)}( \mathbf{z}, \mathbf{\bar{z}})=-f^{(3)}(1- \mathbf{z},1- \mathbf{\bar{z}})=-f^{(3)}\Big(\frac{1}{\mathbf{z}},\frac{1}{\mathbf{\bar{z}}}\Big)\,.
\end{equation}

The function $\mathfrak{f}_s$, which we recall here for convenience,
\begin{equation}
\mathfrak{f}_s(\mathbf{z},\mathbf{\bar{z}}) = \frac{1}{3} \frac{1}{\mathbf{z}-\mathbf{\bar{z}}} \left( f^{(3)} + \Big( \log \mathbf{u}  + \log \frac{\mathbf{u}}{\mathbf{v}} \Big)\phi^{(1)}\right)
\end{equation}
is nothing but the ``$s$-channel version'' of $f^{(3)}/(\mathbf{z}-\mathbf{\bar{z}})$. In fact, it only preserves a $1\rightarrow 2$ crossing transformation:
\begin{equation}
\mathfrak{f}_s(\mathbf{z},\mathbf{\bar{z}})= \frac{1}{\mathbf{v}^2}  \mathfrak{f}_s\left(\frac{\mathbf{z}}{1-\mathbf{z}},\frac{\mathbf{\bar{z}}}{1-\mathbf{\bar{z}}} \right) \,,
\end{equation}
and we have
\begin{equation}
\mathfrak{f}_s(\mathbf{z},\mathbf{\bar{z}})+\mathfrak{f}_t(\mathbf{z},\mathbf{\bar{z}})+ \mathfrak{f}_u(\mathbf{z},\mathbf{\bar{z}}) = \frac{1}{\mathbf{z}-\mathbf{\bar{z}}}  f^{(3)}(\mathbf{z},\mathbf{\bar{z}}) \,,
\end{equation}
where $\mathfrak{f}_t,\mathfrak{f}_u$ are the crossing versions defined in \eqref{ftfu}.

Finally, note that $\mathfrak{f}_s$ admits the $\log \mathbf{u}$ stratification
\begin{equation}
\mathfrak{f}_s(\mathbf{z},\mathbf{\bar{z}}) =\frac{1}{\mathbf{z}-\mathbf{\bar{z}}}\log \frac{1-\mathbf{z}}{1-\mathbf{\bar{z}}} \times \log \mathbf{u} \log \frac{\mathbf{u}}{\mathbf{v}}+\cdots 
\end{equation}
where the dots do not contain $\log^2 \mathbf{u}$, which implies
\begin{equation}
\mathfrak{f}_s\Big|_{\log^2 \mathbf{u}}= \frac{\phi^{(1)}}{\mathbf{z}-\mathbf{\bar{z}}}\Big|_{\log \mathbf{u}} =\frac{1}{\mathbf{z}-\mathbf{\bar{z}}}\log \frac{1-\mathbf{z}}{1-\mathbf{\bar{z}}}\,.
\end{equation}
This, together with the property that
\begin{equation}
\mathfrak{f}_t(\mathbf{z},\mathbf{\bar{z}})\Big|_{\log^2 \mathbf{u}}=\mathfrak{f}_u(\mathbf{z},\mathbf{\bar{z}})\Big|_{\log^2 \mathbf{u}}=0\,,
\end{equation}
imply that the uplift \eqref{generatorcorrelatoroneloopuplift} manifestly reproduces the trascendental weight 2 part of the 1d correlators given by \eqref{generatorcorrelatoroneloop}.

The function $f^{(3)}$ is not new in this context, and it appears in the position-space representation of correlators with a weakly-coupled dual AdS description \cite{Drummond:2019hel,Huang:2021xws,Drummond:2022dxw,Huang:2023oxf,Paul:2023zyr}. In particular, this function appears whenever there is a logarithmic divergence in the associated flat-space amplitude. In fact, an important feature of this function is that it contains the letter $\mathbf{z}-\mathbf{\bar{z}}$ and, as pointed out in \cite{Drummond:2022dxw} (see also \cite{Paul:2023zyr}), the presence of this letter signals a logarithmic divergence $\sim \log (\mathbf{z}-\mathbf{\bar{z}})$ in the bulk-point limit. This is indeed the case for a 4d one-loop flat-space amplitude, which the AdS$_2\times$S$^2$ correlator should approach to in the bulk-point limit.

\section{Sphere integration} \label{sphere}

The sphere integration of the one-loop bubble diagram~\eqref{ks} can be performed explicitly for all charges and for generic dimensions. This is far more general than we need but we give it in full generality as it may be useful in the future. 
We start by expanding all the propagators ${\bf{x}}^2_{ij}=-2x_{i}{\cdot} x_{j}-2y_{i}{\cdot} y_{j}$ in the $y,\hat y$ coordinates:
\begin{align}\label{sphereinteg}
  	&\mathcal{K}_s =  \int d^{d+1}\hat{x}_r d^{d+1}\hat{x}_l \sum_{p_i,k} I_{p_i;k}(y_i) \times  \frac{(2d)_{k}}{k!}\left(\prod_i \frac{(d)_{p_i}}{p_i!} \right) \notag  \\
 & \frac{2^{-6d-2\Sigma_p}}{(x_{1}{\cdot}\hat x_l)^{p_1+d}(x_{2}{\cdot}\hat x_l)^{p_2+d}(\hat x_{l}{\cdot}\hat x_r)^{2d+k}(x_{3}{\cdot}\hat x_r)^{p_3+d}(x_{4}{\cdot}\hat x_r)^{p_4+d}} 
 \end{align}
where
\begin{align}
&I_{p_i;k}(y_i)= 2^{2\Sigma_p} \int d^{d+1}\hat{y}_l d^{d+1}\hat{y}_r (y_{1}{\cdot}\hat y_l)^{p_1}(y_{2}{\cdot}\hat y_l)^{p_2}(\hat y_{l}{\cdot}\hat y_r)^k \notag \\
& 	 \,\qquad \qquad \qquad \qquad \qquad \qquad  \times(y_{3}{\cdot}\hat y_r)^{p_3}(y_{4}{\cdot}\hat y_r)^{p_4}\ 
\label{I}
\end{align}
and $\Sigma_p=(p_1{+}p_2{+}p_3{+}p_4)/2$.
We will now do the $\hat y_{l,r}$ integrations. First consider the three-point function corresponding to the integration over $y_l$
\begin{align}
	J&=\int d^{d+1}y_l\, (y_1 {\cdot} \hat{y}_l)^{p_1}(y_2 {\cdot} \hat{y}_l)^{p_2}(\hat{y}_l {\cdot}\hat{y}_r)^k\notag
	\\&=y_1^{A_1} {\dots} y_1^{A_{p_1}}y_2^{B_1} {\dots} y_2^{B_{p_2}} \hat{y}_r^{E_1} {\dots} \hat{y}_r^{E_k}\int d^{d+1}\hat{y}_l \hat{y}_{l}^{A_1} {\dots} \hat{y}_{l}^{E_{k}}\,. \label{J}
\end{align}
The integral of the product of $2L$ $y_l$'s  gives a factor ${\mathcal N}_{L}$ times a product of symmetrised delta functions 
\begin{align}
	\int d^{d+1}\hat{y}_l \hat{y}_{l}^{A_1} \dots \hat{y}_{l}^{A_{2L}}= {\mathcal N}_{L} \delta^{(A_1A_2} \dots \delta^{A_{2L-1}A_{2L})}
\end{align}
and zero if there are an odd number of $\hat{y}_l$'s (see e.g. \cite{Chen:2020ipe})
with
\begin{align}
	{\mathcal N}_L={\text{vol}}(S^{d+1}) \frac{(2L)!}{2^{2L}L!(d/2+1)_L}\ .
\end{align}
We need to consider then the product of $(p_1+p_2+k)/2$ Kronecker-delta's with indices completely symmetrised: 
\begin{align}
	\delta^{(A_1A_2} \dots \delta^{A_{p_1-1}A_{p_1}} \delta^{B_{1}B_2}\dots \delta^{B_{p_2-1}A_{p_2}}\delta^{E_{1}E_2}\dots \delta^{E_{k-1}E_k)}\label{deltas}
\end{align}
and count the proportion of terms which have the schematic form:
\begin{align}
	(\delta^{AA})^a (\delta^{BB})^b(\delta^{EE})^e(\delta^{AB})^\epsilon(\delta^{AE})^\alpha(\delta^{BE})^\beta
\end{align}
with
\begin{align}
	p_1=2a+\alpha+\epsilon \qquad 	p_2=2b+\beta+\epsilon \qquad 	k=2e+\alpha+\beta\ . \label{schemdelta}
\end{align}
Counting these is equivalent to counting the number of inequivalent graphs of the form
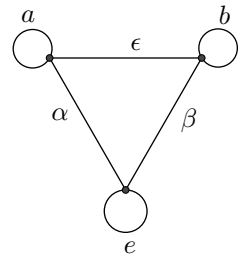
\begin{center}
	\begin{tikzpicture}[scale=0.125]
\draw [line width=.5pt] (-8,6)-- (0,-8);
\draw [line width=.5pt] (0,-8)-- (8,6);
\draw [line width=.5pt] (8,6)-- (-8,6);
\draw [line width=.5pt] (-9.789603839915454,6.9818130617006275) circle (2.041234624428583cm);
\draw [line width=.5pt] (9.726393094402392,7.059111031443233) circle (2.0253763337525728cm);
\draw [line width=.5pt] (0.0004521980997308963,-10.248602953934865) circle (2.248602999403791cm);
\draw (-0.5,9) node[anchor=north west] {$\epsilon$};
\draw (4.8,1.65) node[anchor=north west] {$\beta$};
\draw (-8.8,1.65) node[anchor=north west] {$\alpha$};
\draw (-12,12) node[anchor=north west] {$a$};
\draw (12,12.5) node[anchor=north east] {$b$};
\draw (-1.1726994622587217,-12.5) node[anchor=north west] {$e$};

\draw [fill=uuuuuu] (-8,6) circle (10pt);
\draw [fill=uuuuuu] (8,6) circle (10pt);
\draw [fill=uuuuuu] (0,-8) circle (10pt);
\end{tikzpicture}
\end{center}
where each line represents a number of parallel edges given by the associated label. 
The number of such inequivalent graphs is
\begin{align}
	\frac{p_1!p_2!k!}{a!b!e!\alpha!\beta!\epsilon!2^{a+b+e}}
\end{align}
whereas the total number of inequivalent terms of the form~\eqref{deltas} is $(p_1+p_2+k-1)!!$.

With this information we can obtain  the integral $J$ in~\eqref{J}.  Terms from~\eqref{schemdelta} with $a>0$ or $b>0$ produce $y_1 {\cdot} y_1=0$ or $y_2 {\cdot} y_2=0$ in~\eqref{J} thus only terms with $a=b=0$ survive. Thus
\begin{align}
J= & {\mathcal N}_{(p_1+p_2+k)/2} \sum_e \Big(	(y_1.y_2)^\epsilon  (y_1.\hat y_r)^\alpha (y_2.\hat y_r)^\beta\, \notag \\
&  \frac{p_1!p_2!k!}{e!\alpha!\beta!\epsilon! 2^{e} (p_1{+}p_2{+}k{-}1)!! } \Big)
\end{align}
with
\begin{align}
& 	\alpha =\frac{1}{2} \left({-}2 e{+}k{+}p_1{-}p_2\right)  \notag \\ 
& \beta =\frac{1}{2} \left({-}2e{+}k{-}p_1{+}p_2\right) \notag , \\
& \epsilon =\frac{1}{2} \left(2 e{-}k{+}p_1{+}p_2\right)\,.
\end{align}
Inserting this into \eqref{I} then gives the exchange diagram as a sum of contact sphere diagrams $B_{p_1p_2p_3p_4}(y_i)$ (similarly to what happens for AdS exchange Witten diagrams):
 \begin{align}
  I_{p_i,k}= {\mathcal N}_{\frac{p_1+p_2+k}2} \sum_e & \Big( \frac{p_1!p_2!k!}{e!\alpha!\beta!\epsilon!(p_1{+}p_2{+}k{-}1)!! 2^{e}}  \notag  \\
 	& \times \frac{B_{\alpha \beta p_3 p_4}(y_1,y_2,y_3,y_4)}{2^{\alpha+\beta+p_3+p_4}}\Big)\ \label{IinB}
 \end{align}
where
\begin{align}
	B_{p_1p_2p_3p_4}= 2^{2 \Sigma_p}\int d^{d+1}\hat y (\hat y.y_1)^{p_1}(\hat y.y_2)^{p_2}(\hat y.y_3)^{p_3}(\hat y.y_4)^{p_4}
\end{align}
is the tree-level function which can be explicitly integrated giving:
\begin{equation}
\begin{split}
B_{p_1p_2p_3p_4}(y_i)=  &  {\text{vol}}(S^{d+1})\,\frac{p_1!p_2!p_3!p_4!(d/2)!(-1)^{\Sigma_p}}{((p_1{+}p_2{+}p_3{+}p_4{+}d)/2)!} \\
& \times \sum_{\{ d_{ij}\}}\prod_{i<j} \frac{(y_{ij}^2)^{d_{ij}}}{ d_{ij}!} .\label{B}
\end{split}
\end{equation} 
with
\begin{equation}
\sum_{i}  d_{ij} =p_j\,, \quad  d_{ii}=0\,, \quad  d_{ij}= d_{ji}\ .
\end{equation}
We can thus obtain an explicit expression for the exchange diagram $I$ by inserting this into~\eqref{IinB}.  
 Since it is $B_{\alpha \beta p_3 p_4}$ rather than $B_{p_1p_2p_3p_4}$ we replace $d_{12} \rightarrow d_{12}-\epsilon$ in~\eqref{B}. After some simplification and performing the sum over $e$ we get
 \begin{align}
I_{p_i;k}= (-2)^{\Sigma_p+k}	\sum_{\{ d_{ij}\}} c_{p_i;k}(d_{ij}) \prod_{i<j} (y_{ij}^2)^{d_{ij}}\label{I1}
 \end{align}
with
 \begin{align}
 c_{p_i;k}(d_{ij}) = & 
 \frac{{\text{vol}}(S^{d{+}1})^2  p_1! p_2! p_3! p_4! (d!!)^2 }{d_{12}! d_{13}! d_{14}! d_{23}! d_{24}!d_{34}!  \left(\frac{d}{2}
	{+} \Sigma _p\right)! } \times  \notag \\
	&  \frac{k! \left(\frac{1}{2} \left(k{-}d_{13}{-}d_{14}{-}d_{23}{-}d_{24}\right){+}1\right)_{\Sigma_p+d/2} }{\left(d{+}k{+}p_1{+}p_2\right)!! \left(d{+}k{+}p_3{+}p_4\right)!!}\,.\label{c}
 \end{align}
Plugging everything in \eqref{sphereinteg} after some work one can do the sum over $k$ explicitly. The result is given in terms of a single ${}_3F_2$ hypergeometric function:
\begin{align}\label{Bone}
& \mathcal{K}_s  =\sum_{d_{ij}=0}^\infty \left(\prod_{ i<j} \frac{(y_{ij}^2)^{d_{ij}}}{d_{ij}!}\right)	\frac{(\prod_i (d)_{p_i})(2d)_{d_4}(d/2)!^2 }{2^{d_{12}+d_{34}+d_4-6d+\Sigma_p}}\times \notag\\
& 
 \int \frac{d^{d{+}1}\hat{x}_l d^{d{+}1}\hat{x}_r \,\,
  {}_3F_2 \Big(a_1,a_2,a_3;a_4,a_5;1/(\hat x_l{\cdot}\hat x_r)^2\Big) }{(x_{1}{\cdot}\hat x_l)^{p_1{+}d}(x_{2}{\cdot}\hat x_l)^{p_2{+}d}(\hat x_{l}{\cdot}\hat x_r)^{2d{+}d_4}(x_{3}{\cdot}\hat x_r)^{p_3{+}d}(x_{4}{\cdot}\hat x_r)^{p_4{+}d}}
\end{align}
where 
\begin{align*}
& a_1= \frac{2d+d_4}{2}\,, \\
& a_2= a_1+1/2 \,,\\
&  a_3=\frac{2{+}d{+}2d_{12}{+}2d_{34}{+}2d_4}2\,, \\
&  a_4=\frac{2{+}d{+} 2d_{12} {+} 2d_4}2\,, \\
&  a_5= \frac{2 {+} d {+} 2d_{34} {+} 2d_4}2\,, \\
\end{align*}
and $d_4=d_{13}+d_{14}+d_{34}+d_{24}$. Note that in the case of $p_i=0, d=1$, the hypergeometric is $$\frac{1}{(\hat x_{l}{\cdot}\hat x_r)^2}\,{}_3F_2\Big(1,\frac{3}{2},\frac{3}{2};\frac{3}{2},\frac{3}{2},\frac{1}{(\hat x_{l}{\cdot}\hat x_r)^2}\Big)=\frac1{(\hat x_l{\cdot}\hat x_r)^2-1}\ .$$ This gives the result quoted in the main text, see formula \eqref{Ksp=1}.

\section{AdS Integral} \label{adsdetails}
In this appendix we give some more details on the AdS integration of the bubble diagram. 
We will restrict to $d=1$, which is the theory we are interested in, and focus on the simplest case, i.e. $p_i=0$.
Nicely, plugging $d=1$ in \eqref{Bone}  and projecting onto the $y_i=0$ component, the hypergeometric simplifies to a geometric function. In the end, we are left with the following double integral:
\begin{equation}
\mathcal{K}_{s}\big|_{\substack{d=1\\y_{i}=0}}=\!\!\int\!\!\frac{d^{2}\hat{x}_{l}d^{2}\hat{x}_{r}\times 2^{-6}}{(x_{1}{\cdot}x_l)(x_{2}{\cdot}x_l)(x_{3}{\cdot}x_r)(x_{4}{\cdot}x_r)\big((\hat{x}_{l}{\cdot}\hat{x}_{r})^{2}{-}1\big)}\,,
\end{equation}
and its crossing orientations \footnote{From now on, we will always consider $\mathcal{K}_s$ evaluated at $d=y_i+1=1$ and hence drop the notation $|_{d=1, y_i=0}$.}.
This integral can be evaluated following the logic outlined in \cite{Heckelbacher:2022fbx}. The idea is to turn the AdS integral into a flat-space integral and evaluate it by borrowing known flat-space techniques.

To this extent, let us map euclidean AdS to the upper half plane:
\begin{equation}
\mathcal{C}_2 = \{Z= (x,z),\, x\in \mathbb{R} ,\, z>0 \}.
\end{equation}
The bulk coordinates are then parametrised as follows
\begin{align}
\hat{x}& =\biggl(\hat{x}^{(-1)},\hat{x}^{(0)},\hat{x}^{(1)}\biggr)= \notag \\
& = \frac{1}{\sqrt{2} z}\left(1+\frac{1}{2}|Z|^2, \sqrt{2} x, 1-\frac{1}{2}|Z|^2 \right),
\end{align}
with $\hat{x} {\cdot}\hat{x}= -\big(\hat{x}^{(-1)}\big)^2+\big(\hat{x}^{(0)}\big)^2+\big(\hat{x}^{(1)}\big)^2=-1$,
and we have defined the euclidean distance,
\begin{equation}
|Z|^2 \equiv  x^2+z^2.
\end{equation}
Analogously, the boundary coordinates are parametrised in terms of the 1d coordinate $x_i$ as
\begin{align}\label{embedding}
x_i^A & =\biggl(x_i^{(-1)},x_i^{(0)},x_i^{(1)}\biggr) \notag \\
& =\frac{1}{\sqrt{2}}\left(1+\frac{1}{2}x_i^2, \sqrt{2} x_i, 1-\frac{1}{2}x_i^2 \right),
\end{align}
with ${x}_i {\cdot} {x}_i=0$.

In these coordinates, we have
\begin{align}
& \hat{x}_l\cdot \hat{x}_r= -1-\frac{|Z_l-Z_r|^2}{2(p\cdot Z_l)(p\cdot Z_r)}\,,\\
& x_i \cdot \hat{x}_l= -\frac{|Z_i-Z_l|^2}{2(p\cdot Z_l)},\qquad \qquad  Z_i=(x_i,0)\,,
\end{align}
where we introduced a constant vector $p \in \mathbb{R}^2$ \footnote{Note that this is the scalar product in $\mathbb{R}^2$ equipped with the euclidean metric.}:
\begin{equation}
p\cdot Z_{l,r}=z_{l,r}, \qquad p=(0,1)\,,
\end{equation}
and we also embedded the boundary coordinates in $\mathbb{R}^2$ so that:
\begin{equation}
|Z_l-Z_i|^2 = (x_l-x_i)^2+z_l^2, \qquad Z_i=(x_i,0).
\end{equation}
Now, note that the effective propagator satisfies the following identity:
\begin{equation}\label{identitybtob}
\begin{split}
& \frac{2}{(\hat{x}_l{\cdot} \hat{x}_r)^2-1}=\left(\frac{1}{\hat{x}_l{\cdot} \hat{x}_r-1} -\frac{1}{\hat{x}_l{\cdot} \hat{x}_r+1} \right)=\\
&=\frac{2z_l z_r}{(z_l-z_r)^2+(x_l-x_r)^2}-\frac{2z_l z_r}{(z_l+z_r)^2+(x_l-x_r)^2} \\
&= \frac{2(p\cdot Z_l)( p\cdot Z_r)}{|Z_l-Z_r|^2}+\frac{2(p\cdot Z_l)( p\cdot \sigma(Z_r))}{|Z_l-\sigma(Z_r)|^2}
\end{split}
\end{equation}
where $\sigma(Z)$ is the antipodal map
\begin{equation}
\sigma(x,z):= (x,-z).
\end{equation}
The trick is then to use the identity \eqref{identitybtob} to split the integral and then rewrite it as a single integral with the domain of integration extended to the full plane, so that it effectively becomes a flat-space integral:
\begin{align}
 & \mathcal{K}_s  = \frac{1}{2} \int_{\mathbb{R}^{4}}\frac{d^2 Z_l \, d^2 Z_r}{(p\cdot Z_l)^2(p\cdot Z_r)^2} \times \notag \\
 & \frac{(p\cdot Z_l)^3( p\cdot Z_r)^3}{|Z_l{-}Z_r|^2|Z_l{-}Z_1|^2|Z_l{-}Z_2|^2|Z_r{-}Z_3|^2|Z_r{-}Z_4|^2}\,.
\end{align}

The integral contains divergences and it needs to be regularised. A way to do this is to change the dimension of integration without changing the pre-factor coming from the metric \cite{Heckelbacher:2022fbx}:
\begin{equation}
\frac{d^2 Z}{(p\cdot Z)^2}\quad \rightarrow \quad  \frac{d^D Z}{(p\cdot Z)^2}, \qquad  D=2-2\epsilon.
\end{equation}
The regularised integral becomes
\begin{align}
& \mathcal{K}_s^{(\epsilon)} \equiv \frac{1}{2} \int_{\mathbb{R}^{2D}}\frac{d^D Z_l \, d^D Z_r}{(p\cdot Z_l)^2(p\cdot Z_r)^2}   \mathcal{I}
\end{align}
where the integrand $\mathcal{I}$ reads
\begin{align}
& \mathcal{I}\equiv\frac{(p{\cdot} Z_l)^3( p{\cdot} Z_r)^3}{|Z_l{-}Z_r|^2|Z_l{-}Z_1|^2|Z_l{-}Z_2|^2|Z_r{-}Z_3|^2|Z_r{-}Z_4|^2} \,.
\end{align}
Here, $D=2-2\epsilon$ and the superscript $(\epsilon)$ emphasises that the integral has been regularised. 

To simplify the evaluation of the integral, we will take advantage of conformal symmetry to recast it in terms of conformal cross ratios.
In doing so, we need to be careful as the dimensional regularisation scheme we are using breaks conformal invariance, as a consequence of the fact that we changed the dimension without changing the measure factor.
Let us first take care of the integrand and then of the measure.
Let us shift every boundary point by $Z_4$ and then invert every bulk point. The inversion only changes the bulk-to-boundary propagator which maps to:
\begin{equation}
\frac{p\cdot Z_l}{|Z_l{-}Z_i|^2} \rightarrow \frac{p\cdot Z_l'}{|Z_l'{-}Z_i|^2}= \frac{1}{|Z_i|^2}  \frac{p\cdot Z_l}{|Z_l{-}\frac{Z_i}{|Z_i|^2}|^2}
\end{equation} 
Then, we shift every bulk point by $Z_{l,r} \rightarrow Z_{l,r} + \tilde{Z}_{14}$ where we defined $\tilde{Z}_{ij}=\frac{Z_{ij}}{Z_{ij}^2}$, $Z_{ij}=Z_i-Z_j$ and then rescale every bulk point by $Z_{l,r} \rightarrow \left( \tilde{Z}_{34}-\tilde{Z}_{14} \right)Z_{l,r}$ \footnote{Note that in the case of one-dimensional boundary we could also simplify $\frac{Z_{ij}}{Z_{ij}^2}=\frac{1}{Z_{ij}}$. It is however convenient to keep the dimension generic.}.
Thus, after these transformations the integrand reads \footnote{Note that $Z_{ij}^2=x_{ij}^2$.}
\begin{align}
\mathcal{I}= & \frac{1}{x_{13}^2 x_{24}^2}\frac{(p\cdot Z_l)^3( p\cdot Z_r)^3}{|Z_l{-}Z_r|^2|Z_l|^2} \notag \\
& \times \frac{1}{| Z_l{-}\frac{\tilde{Z}_{24}{-} \tilde{Z}_{14}}{|\tilde{Z}_{34}{-} \tilde{Z}_{14}|^2}|^2 | Z_r{-}\frac{\tilde{Z}_{34}{-} \tilde{Z}_{14} }{|\tilde{Z}_{34}{-} \tilde{Z}_{14}|}|^2}  .
\end{align}
We can finally use the conformal group acting on the boundary to map $Z_3$ to infinity, $Z_4$ to the origin, $Z_1=(-1,0)$ and parametrise $Z_2$ with the conformal ratio $x$ \footnote{In $d>1$, $Z_2$ is parametrised by two cross ratios:
\begin{equation}\label{Z22cross}
Z_2= \Big(\frac{\mathbf{z}+\mathbf{\bar{z}}-2}{2(1-\mathbf{z})(1-\mathbf{\bar{z}})},\frac{\mathbf{z}-\mathbf{\bar{z}}}{2i(1-\mathbf{z})(1-\mathbf{\bar{z}})},0,\ldots,0 \Big)^T\,.
\end{equation}}:
\begin{equation}
Z_2 = \left(-\frac{1}{1-x},0 \right)\,,\qquad \qquad x=\frac{x_{12}x_{34}}{x_{13}x_{24}}.
\end{equation}
This gives:
\begin{equation}
\mathcal{I}=\frac{1}{x_{13}^2 x_{24}^2}\frac{(p\cdot Z_l)^3( p\cdot Z_r)^3}{|Z_l{-}Z_r|^2|Z_l|^2}  \frac{1}{| Z_l{-}m_x|^2} \frac{1}{| Z_r{-}m_1|^2},
\end{equation}
with $m_1=(1,0)$, $m_x=(x,0)$.

On the other hand, the measure, after all the above conformal transformations, maps to 
\begin{align}
& \int_{\mathbb{R}^{2D}}\frac{d^D Z_l \, d^D Z_r}{(p\cdot Z_l)^2(p\cdot Z_r)^2} \rightarrow \notag \\
&  \int_{\mathbb{R}^{2D}}\frac{d^D Z_l \, d^D Z_r}{(p\cdot Z_l)^2(p\cdot Z_r)^2}\frac{1}{| Z_l {-} m_1 |^{2(D-2)} | Z_r {-} m_1 |^{2(D-2)}}\,.
\end{align}
Putting all together we arrive at the following integral
\begin{align}
\mathcal{K}_s^{(\epsilon)} =& \frac{1}{2 x_{13}^2 x_{24}^2}  \int_{\mathbb{R}^{2D}}d^D Z_l \, d^D Z_r \notag \\
& \times   \frac{(p\cdot Z_l)( p\cdot Z_r)|Z_l{-}m_1|^{4\epsilon}}{|Z_l-Z_r|^2|Z_l|^2|Z_l{-}m_{x}|^2|Z_r{-}m_1|^{2-4\epsilon}}.
\end{align}
This regularised integral can be analytically evaluated with standard flat-space techniques.
In short, one first introduces Feynman parameters, which allow to perform the spacetime integration. Then, the remaining parametric integral can be computed using e.g. the \texttt{HyperInt} package \cite{Panzer:2014caa}.

\vspace{0.5cm}

{\bf Feynman parameters and analytic evaluation of the integral}.
Let us recall a few useful formulae for the evaluation of the AdS integral. We begin with the familiar Feynman parametrisation:
\begin{align}\label{Feynmanpar}
& \frac{1}{A_1^{\alpha_1}\cdots A_n^{\alpha_n}} =\notag \\
& = \frac{\Gamma(\alpha_1+\cdots+\alpha_n)}{\Gamma(\alpha_1)\cdots \Gamma(\alpha_n) } \int_0^1  \frac{\prod_{i=1}^n d l_i  l_i^{\alpha_i-1} \, \delta \big(1-\sum_{i=1}^n l_i \big)}{(\sum_{i=1}^n l_i A_i )^{\sum_{i=1}^n \alpha_i }}.
\end{align}
Let us also recall the \emph{one-loop master formula} (see e.g. \cite{Weinzierl:2022eaz})
\begin{equation}
\int \frac{d^D k}{i\pi^{\frac{D}{2}}} \frac{(-k^2)^a}{(-U k^2 + V)^b}=\frac{\Gamma(\frac{D}{2}+a)\Gamma(b-\frac{D}{2}-a)}{\Gamma(\frac{D}{2})\Gamma(b)} \frac{U^{-\frac{D}{2}-a}}{V^{b-\frac{D}{2}-a}},
\end{equation}
and a very useful AdS-variation of it:
\begin{equation}\label{onelmasterintegravariation}
\begin{split}
& \frac{1}{\pi^{d+1}} \int d^d z \, d z_0\, d^d w \, d w_0 \frac{ (w_0^2)^{a_1} (z_0^2)^{a_2}}{(c_1 w^2+ c_2 w_0^2+c_3 z^2+ c_4 z_0^2+c_5)^b}\\
 & =\, \frac{1}{\pi \Gamma(b)} \Gamma \Big(a_1+\frac{1}{2} \Big)\Gamma \Big(a_2+\frac{1}{2}\Big)\Gamma \Big(b-d-1-a_1-a_2\Big) \times\\
&  \qquad \frac{c_1^{-\frac{d}{2}} c_2^{-\frac{1}{2}-a_1}c_3^{-\frac{d}{2}}c_4^{-\frac{1}{2}-a_2}}{c_5^{b-d-1-a_1-a_2}}.
\end{split}
\end{equation}
This is all we need to turn the AdS integrals involved in our one-loop bubble diagram into integrals over Feynman parameters. 
Let us first rewrite the integrand using the Feynman parametrisation \eqref{Feynmanpar}:
\begin{equation}
\begin{split}
& \frac{(p\cdot Z_l)( p\cdot Z_r)}{|Z_r{-} m_1|^{2{-} 4\epsilon}|Z_l|^2|Z_l{-} m_1|^{{-} 4\epsilon} |Z_l{-} m_{x}|^2  |Z_l{-} Z_r|^2} \\
 & = \frac{\Gamma(4{-} 4\epsilon)}{\Gamma(1-2\epsilon)\Gamma(-2\epsilon)} \int \left(\prod_{i=1}^5 dl_i\right) l_1^{-2\epsilon}  l_3^{-1-2\epsilon} \delta \big(1{-} \sum_{i} l_i \big) \times \\
&  \qquad \frac{(p\cdot Z_l)( p\cdot Z_r)}{(\beta_1|Z_r {-}  Z' |^2 +\beta_2 |Z_l {-} Z'' |^2+\beta_3 )^{4-4\epsilon}}\,,
\end{split}
\end{equation}
where $Z',Z''$ are two-component vectors, $\beta_i$ are constants and they do not depend on $Z_r$:
\begin{align}
& \beta_1 = l_1+l_5 \,, \notag \\
& \beta_2 = - \frac{(l_2+l_3+l_4)l_5+(l_2+l_3+l_4+l_5)l_1}{l_1+l_5}\,, \notag \\
& \beta_3= \frac{l_1(l_2(l_3 + l_4 x^2+l_5  ) +l_3 l_4(1-x )^2+l_4 l_5(1-x)^2 )}{(l_1+l_5)\beta_2} +  \notag  \,\\
& \qquad   + \frac{ l_5(  l_2(l_3+l_4 x^2) +l_3 l_4(1-x)^2)}{(l_1+l_5)\beta_2} \notag \,, \\
& Z' =  \frac{l_1 m_1 +l_5 Z_l}{l_1+l_5}\,,\notag \\
& Z'' =  \frac{l_1(l_3 m_1 + l_4 m_x +l_5 m_1 )+l_5(l_3 m_1 + l_4 m_x ) }{(l_1+l_5) \beta_2 } \,,
\end{align}
where we used the fact that $||m_1||=1,||m_x||=x$.
Now, since this expression will be integrated in $Z_r$, we can shift $Z_r\rightarrow Z_r +Z'$. Under this shift, the numerator maps to \footnote{Note that $p\cdot m_x=p\cdot m_1=0$.}
\begin{equation}
(p\cdot Z_l)( p\cdot Z_r) \rightarrow (p\cdot Z_l)( p\cdot Z_r)+ \frac{l_5}{l_1+l_5}(p\cdot Z_l)^2.
\end{equation} 
The first integrand is odd in $Z_R$, thus it vanishes upon integration.
We are thus left with the following integrand
\begin{equation}
\begin{split}
 &  \frac{\Gamma(4-4\epsilon)}{\Gamma(1-2\epsilon)\Gamma(-2\epsilon)} \int \prod_{i=1}^5 dl_i \,\,l_1^{-2\epsilon}  l_3^{-1-2\epsilon} \delta \big(1-\sum_{i} l_i \big) \times \\
 & \qquad \frac{l_5}{l_1+l_5} \frac{( p\cdot Z_l)^2}{(\beta_1|Z_r|^2 +\beta_2 |Z_l- Z'' |^2+\beta_3 )^{4-4\epsilon}}.
\end{split}
\end{equation}
After performing the shift $Z_l\rightarrow Z_l +Z''$, the integrand takes the form \eqref{onelmasterintegravariation} with $c_1=c_2=\beta_1, c_3=c_4=\beta_2, c_5=\beta_3,a_1=0,a_2=1,b=4-4\epsilon$ and $d=1-2\epsilon$.
The double spacetime integration in $Z_l,Z_r$ thus gives
\begin{equation}
\begin{split}
&  \mathcal{K}_s^{(\epsilon)} =\frac{1}{2 x_{13}^2 x_{24}^2}\frac{\Gamma(4-4\epsilon)}{\Gamma(1-2\epsilon)\Gamma(-2\epsilon)} \int_{\mathbb{R}^{2D}}d^{2D} Z_l d^{2D}  Z_r \times \\
 &  \int \prod_{i=1}^5 dl_i\,\, \frac{l_1^{-2\epsilon}  l_3^{-1-2\epsilon}l_5 \delta \big(1-\sum_{i} l_i \big) ( p\cdot Z_l)^2}{(l_1+l_5)(\beta_1|Z_r|^2 +\beta_2 |Z_l |^2+\beta_3 )^{4-4\epsilon}} \\
& =  \frac{1}{4 x_{13}^2 x_{24}^2} \frac{\pi^{2-2\epsilon}}{ \Gamma(-2\epsilon)}  \int \prod_{i=1}^5 dl_i   \frac{ l_1^{-2\epsilon}l_3^{-1-2\epsilon}l_5 ((l_1{+}l_5)\beta_2)^{-2+\epsilon}}{ \beta_3^{1-2\epsilon}}   \,,
\end{split}
\end{equation}
which can be evaluted with the \texttt{HyperInt} package \cite{Panzer:2014caa}. In fact, this job has actually already been done in \cite{Heckelbacher:2022fbx}: upon inspection, it turns out that the parametric integral is precisely equal to (the diagonal limit of) their expressions (C.4)-(C.7). The result can be written as
\begin{align}
\mathcal{K}_s^{(\epsilon)} = & \frac{\pi^{2-2\epsilon} e^{-2\gamma_{\epsilon}}}{16 \mathbf{x}_{13}^2  \mathbf{x}_{24}^2} \times \notag \\
& \left.\left( -\frac{2}{\epsilon} \bar{D}_{1111} +  \mathfrak{f}_s - 2 \log \mathbf{v} \bar{D}_{1111}  +\mathcal{O}(\epsilon) \right)\right|_{y_i=0}
\end{align}
where  $\bar{D}_{1111} = \frac{\phi^{(1)}}{\mathbf{z}-\mathbf{\bar{z}}}$, and similarly for the other orientations.
Let us conclude with two comments. First, while the result is really a function of the 1d cross ratio $x$, we have highlighted the fact it has a natural interpretation as the $y_i=0$ component of a higher dimensional function \footnote{This expression is obtained by simply parametrising $Z_2$ with the two cross ratios $\mathbf{z}, \bar{\mathbf{z}}$, as in equation \eqref{Z22cross}.} .
Second, note that, because of the term $\log \mathbf{v} \bar{D}_{1111}$, the expression is not invariant under $1\rightarrow 2$ crossing. This had to be expected as the regularisation scheme breaks conformal invariance. However, as explained in \cite{Heckelbacher:2022fbx}, the non-crossing symmetric part is exactly the order $\epsilon$ of the tree-level contact diagram in $D=2-4\epsilon$ and thus it can be removed upon renormalising the bare four-point coupling, yielding a finite, crossing-symmetric, and conformally-covariant function.
In other words, the one-loop bubble diagram can be rewritten as
\begin{equation}
\mathcal{K}_s^{(\epsilon)} = -\left. \frac{\pi^{2-2\epsilon}e^{-2\gamma_{\epsilon}}}{16 \mathbf{x}_{13}^2 \mathbf{x}_{24}^2}\left(\frac{2}{\epsilon}  \bar{D}_{1111}^{(\epsilon)}+ \mathfrak{f}_s+\mathcal{O}(\epsilon)  \right)\right|_{y_i=0}\,,
\end{equation}
where $\bar{D}_{1111}^{(\epsilon)}$ is precisely the $\epsilon$-expansion of the four-point contact Witten diagram,
\begin{equation}
\bar{D}_{1111}^{(\epsilon)} = \bar{D}_{1111} - \epsilon \left( \mathfrak{f}_s  +\log  \mathbf{v} \bar{D}_{1111}  \right)+\mathcal{O}(\epsilon^2)\,.
\end{equation}

\bibliographystyle{apsrev4-1}

\end{document}